\documentclass[aps,prd
,nofootinbib,showpacs
,floatfix,superscriptaddress,preprintnumbers]{revtex4}
\usepackage{graphicx}
\usepackage{axodraw}
\usepackage{epsf}
\usepackage{epstopdf}
\usepackage{amsmath}
\usepackage{amssymb}
\usepackage{epsfig}
\usepackage{stmaryrd}
\usepackage{pstricks}
\usepackage{pst-coil}
\DeclareMathAlphabet{\mathsc}{OT1}{cmr}{m}{sc}

\newcommand{\beq}{\begin{equation}}
\newcommand{\eeq}{\end{equation}}
\newcommand{\bea}{\begin{eqnarray}}
\newcommand{\eea}{\end{eqnarray}}
\newcommand{\bec}{\begin{center}}
\newcommand{\eec}{\end{center}}
\newcommand{\bei}{\begin{itemize}}
\newcommand{\eei}{\end{itemize}}

\newcommand{\nn}  {\nonumber}

\def\10{$SO(10)$}
\def\21{SU(2) $\otimes$ U(1) }

\def\422{$SU(4) \otimes SU(2) \otimes SU(2)$}
\def\321{SU(3) $\otimes$ SU(2) $\otimes$ U(1)}

\def\lsim{\raise0.3ex\hbox{$\;<$\kern-0.75em\raise-1.1ex\hbox{$\sim\;$}}}
\def\gsim{\raise0.3ex\hbox{$\;>$\kern-0.75em\raise-1.1ex\hbox{$\sim\;$}}}
\def\vev#1{\left\langle #1\right\rangle}
\def\eq#1{eq.~(\ref{#1})}

 \newcommand{\ba}{\begin{array}}
\newcommand{\ea}{\end{array}}
\relax

\def\321{$SU(3)\times SU(2)\times U(1)$}

\begin{document}

\title{Minimal Dynamical Inverse See Saw}

\author{Federica Bazzocchi\\
\vspace{2mm}
\it{Department of Physics and Astronomy, Vrije Universiteit Amsterdam,\\
1081 HV Amsterdam, The Netherlands}}



\begin{abstract}
We present a minimal model in which the Inverse See Saw is realized dynamically. The two unity lepton number breaking term is induced at two-loop level and is naturally around the keV scale, while right-handed neutrinos are at the TeV scale. An interesting  extension of the model  is obtained by gauging $B-L$: in this case  anomaly cancellation has as direct consequence the presence of a sterile neutrino at the MeV scale that may be a good  Dark Matter candidate. Moreover the new gauge boson $Z'$ and the new neutral scalars may have characteristic signatures at LHC.   \end{abstract}


\maketitle

\section{Introduction}

Experimental evidences of --tiny-- neutrino masses\cite{Maltoni:2004ei} have motivated  the development of a plethora of mechanisms that may explain their riseup and their smallness with respect to the other Standard Model (SM) fermion masses. Definitely the best known mechanism is the  See Saw (SS) mechanism,  usually called type I SS\cite{Minkowski:1977sc} that ascribes to a very high new physics scale. Unfortunately, if  nature had chosen type I  SS we would not have any hope to confirm it at  Large Hadron Collider
(LHC) experiments.

Among all  the mechanisms that provide neutrino masses a very interesting  possibility  is  the so called Inverse See Saw (ISS) mechanism\cite{mohapatra:1986bd,gonzalez-garcia:1989rw}.  In this scheme  no new physics above the TeV scale is introduced and  the smallness of neutrino masses is justified by the smallness of a parameter that breaks the lepton number by two unity, namely  $\mu$. In the limit in which this parameter goes to zero  lepton number is restored  and neutrino are massless.  Being the new physics scale around the TeV  this model is quite appealing for LHC searches. Still the community  has always showed a sizable 	skepticism against this mechanism due to the difficulty in justifying the $\mu$ smallness.  Whereas in the original formulation of the model this was ascribed to   superstring inspired $E_6$ scenario\cite{mohapatra:1986bd},   recently an attempt has been done in the supersymmetric ISS\cite{Bazzocchi:2009kc}, where the smallness of $\mu$  was related to vanishing trilinear susy soft terms at the Grand Unified Theory (GUT) scale. Renormalization group equations (RGEs) both induced them and furnished a dynamical mechanism to justify the $\mu$ size. However  this mechanism had to appeal to a string inspired scenario that could provide vanishing trilinear terms at the GUT scale. So far a very appealing picture is the radiative origin of the two unity lepton number breaking parameter as it has been proposed in \cite{Ma:2009gu}:
it is induced at two-loop level, thus explaining its smallness with respect to the electroweak scale (EW). However in \cite{Ma:2009gu}
an  $SO(10)$ inspired  approach is used hence implying the presence of many new degrees of freedom.

Here we use a bottom-top approach building the dynamical ISS step by step   to satisfy the critieria of  \emph{naturalness} and \emph{minimality}:  we reject \emph{ad hoc} fine tuning in the potential parameters and we  look for the minimal set of ingredients needed to allow the mechanism working. For this reason we start dealing with the global lepton number without introducing extra fermions with respect to the usual ISS model but only new scalar fields. From this point of view the model proposed is quite different from \cite{Ma:2009gu}, where the new fermions play a crucial role in the loops that generate the two unity  lepton number breaking terms.

The paper is organized in this way:  the second  section of the paper revises the ISS idea  and explains the main difficulties in generating dynamically the $\mu$ term preserving naturalness. The third section describes the mechanism proposed while the fourth one  sketches an appealing extension obtained by gauging $B-L$. The latter formulation is less minimal  but  more phenomenologically interesting. Moreover it  has the nice feature to have a MeV Dark Matter (DM)  candidate.

\section{Towards a dynamical ISS realization: naturalness problem}
\label{tow}

In this section we briefly review the ISS mechanism and  present  the problems  related to its dynamical version.

The ISS model is realized by adding to the SM field content two kind of sterile fermions, the usual right-handed  neutrino, $\nu^c$, and a new singlet $S$, charged under  lepton number $-1$ and $1$ respectively. The  lagrangian is invariant under the lepton number  except for a very tiny majorana mass term, $\mu$, involving the new singlet $S$. Due to their singlet nature and lepton charges  $\nu^c$ and $S$ share a Dirac mass term, $M$. The   Yukawa lagrangian  relevant  for  neutrino masses is given by
\beq
\label{ISS0}
\mathcal{L}= y_\nu\, L h \nu^c+ M \, \nu^c S + \frac{1}{2}\mu \, S S+ H.c.\,,
\eeq
where $L$ is the $SU(2)$ lepton doublet, $h$ the standard higgs doublet and  for simplicity we consider only one lepton generation.

When the EW  symmetry is broken by the higgs vacuum expectation value (VEV)  $\vev{h}=v_W/\sqrt{2}$, the neutrino Dirac mass term  $m_D=y_\nu v_W/\sqrt{2}$ is generated. In  the basis $(\nu_L,\nu^c,S)$ the neutrino mass matrix  is given by
\beq
\label{mass0}
M_\nu= \left ( \begin{array}{ccc} 0&m_D &0\\ m_D&0 & M\\ 0 &M& \mu \end{array}\right)\,,
\eeq
and a tiny neutrino mass is generated
\beq
m_\nu\sim \mu \frac{m_D^2}{M^2}\,.
\eeq
Clearly since $m_D$ is fixed around the EW scale $\sim 100$ GeV  and $|m_\nu|\leq $ eV, $\mu$ and $M$ are related and 
\beq
M\geq \sqrt{ (\mu/\mbox{keV})}\sqrt{10} \,\mbox{TeV}\,,
\eeq
thus for $\mu\sim \mathcal{O}(\mbox{keV})$ the singlet neutrino mass is around the TeV  scale, making the model more phenomenologically interesting with respect to the high energy SS realizations.

In order to generate dynamically $\mu$ we could think that the lepton number is spontaneously broken by the VEV of a SM singlet $\Delta$  with lepton number $-2$. Since $\mu\sim\vev{\Delta}$  we should furnish an argument to justify   why $\vev{\Delta}\sim $ keV  whereas  the natural scale is the EW one.

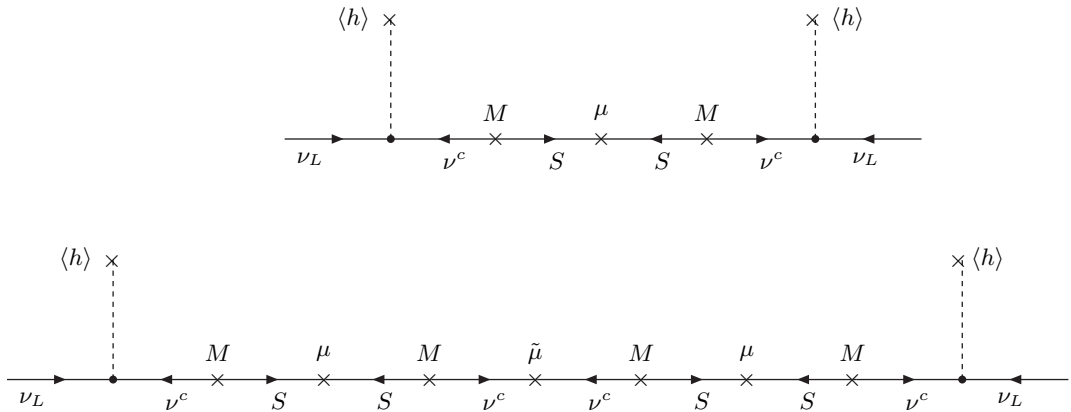
\begin{figure}
\begin{picture}(300,90)(0,0)
\Vertex(90,15){1.5}
\Vertex(250,15){1.5}
 \DashLine(90,60)(90,15){2}
  \DashLine(250,60)(250,15){2}
\Text(90,60)[c]{$\times$}
\Text(250,60)[c]{$\times$}
\Text(170,15)[c]{$\times$}
\Text(130,15)[c]{$\times$}
\Text(210,15)[c]{$\times$}
\Text(130,25)[c]{$M$}
\Text(210,25)[c]{$M$}
\Text(170,25)[c]{$\mu$}
\ArrowLine(210,15)(170,15)
\ArrowLine(210,15)(250,15)
\ArrowLine(290,15)(250,15)
\ArrowLine(130,15)(170,15)
\ArrowLine(130,15)(90,15)
\ArrowLine(50,15)(90,15)
\Text(65,10)[rt]{$\nu_L$} 
\Text(265,10)[l t]{$\nu_L$}
\Text(110,10)[l t]{$\nu^c$}
\Text(150,10)[l t]{$S$}
\Text(190,10)[l t]{$S$}
\Text(230,10)[l t]{$\nu^c$}
\Text(70,65)[l t]{$\vev{h}$}
\Text(257,65)[l t]{$\vev{h}$}
\end{picture}
\begin{picture}(600,90)(0,0)
\Vertex(90,15){1.5}
\Vertex(410,15){1.5}
 \DashLine(90,60)(90,15){2}
  \DashLine(410,60)(410,15){2}
\Text(90,60)[c]{$\times$}
\Text(410,60)[c]{$\times$}
\Text(170,15)[c]{$\times$}
\Text(130,15)[c]{$\times$}
\Text(210,15)[c]{$\times$}
\Text(250,15)[c]{$\times$}
\Text(290,15)[c]{$\times$}
\Text(330,15)[c]{$\times$}
\Text(370,15)[c]{$\times$}
\Text(130,25)[c]{$M$}
\Text(210,25)[c]{$M$}
\Text(290,25)[c]{$M$}
\Text(370,25)[c]{$M$}
\Text(170,25)[c]{$\mu$}
\Text(250,25)[c]{$\tilde{\mu}$}
\Text(330,25)[c]{$\mu$}
\ArrowLine(210,15)(170,15)
\ArrowLine(210,15)(250,15)
\ArrowLine(290,15)(250,15)
\ArrowLine(290,15)(330,15)
\ArrowLine(370,15)(330,15)
\ArrowLine(370,15)(410,15)
\ArrowLine(130,15)(170,15)
\ArrowLine(130,15)(90,15)
\ArrowLine(50,15)(90,15)
\ArrowLine(450,15)(410,15)
\Text(65,10)[rt]{$\nu_L$} 
\Text(435,10)[rt]{$\nu_L$} 
\Text(110,10)[l t]{$\nu^c$}
\Text(150,10)[l t]{$S$}
\Text(190,10)[l t]{$S$}
\Text(230,10)[l t]{$\nu^c$}
\Text(270,10)[l t]{$\nu^c$}
\Text(310,10)[l t]{$S$}
\Text(350,10)[l t]{$S$}
\Text(390,10)[l t]{$\nu^c$}
\Text(70,65)[l t]{$\vev{h}$}
\Text(415,65)[l t]{$\vev{h}$}
\end{picture}
 \caption{The origin of neutrino masses in the ISS model where both  the $\mu$ and $\tilde{\mu}$ terms are present. The contribution proportional to $\tilde{\mu}$ is subleading.} \label{grISSa}
\end{figure}

Indeed if we add at the ISS field content a  SM singlet $\Delta$  with lepton number $-2$ and assume that the lagrangian is  lepton number invariant the Yukawa lagrangian in \eq{ISS0} is replaced by

\beq
\label{ISS1}
\mathcal{L}= y_\nu\, L H \nu^c+ M \, \nu^c S + \frac{1}{2}\,y_S\Delta  \, S S\,+ \frac{1}{2}\,y_{\nu^c} \Delta^\dag  \nu^c \nu^c+H.c.\,,
\eeq
yielding to a neutrino mass matrix
\beq
\label{mass1}
M_\nu= \left ( \begin{array}{ccc} 0&m_D &0\\ m_D&\tilde{\mu} & M\\ 0 &M& \mu \end{array}\right)\,,
\eeq
with $\tilde{\mu}\sim \mu$. However $\tilde{\mu}$ enters in the light neutrino masses only at next to leading order. Using the block diagonalization  method introduced by \cite{Schechter:1980gr}  it is easy to see that the light neutrino mass is still given by 
\beq
m_\nu\sim \mu \frac{m_D^2}{M^2}\,,
\eeq
while the two heavy states have masses
\beq
\pm (M+\frac{m_D^2}{2M})+\frac{1}{2}(\mu+\tilde{\mu})- \mu \frac{m_D^2}{2 M^2}\,.
\eeq
This may be easily understood by looking at the Feynman diagram in fig.~\ref{grISSa} where it is clear that $\tilde{\mu}$ enters only at the second order level.

In the presence of the new singlet $\Delta$ the lepton number--and SM gauge symmetries--scalar invariant potential is given by
\bea
V[h,\Delta]&=& \mu^2_h (h^\dag h)+ \mu^2_\Delta (\Delta^\dag \Delta)+\lambda_h (h^\dag h)^2+ \lambda_\Delta (\Delta^\dag \Delta)^2+ \lambda_{h\Delta} (h^\dag h)(\Delta^\dag \Delta)\,.
\eea

By imposing the vacuum configuration
\beq
\vev{h}=v_W/\sqrt{2} \,,\quad \vev{\Delta}=v_\Delta\,,
\eeq
the minimum equations give:
\bea
\label{vevs}
\frac{v^2_W}{2}= \frac{2 \lambda_\Delta \mu^2_h- \lambda_{h \Delta}\mu^2_\Delta}{\lambda_{h\Delta}^2-4 \lambda_h \lambda_\Delta}&\quad& v_\Delta^2 =\frac{2 \lambda_h \mu^2_\Delta-  \lambda_{h \Delta} \mu^2_h}{\lambda_{h\Delta}^2-4 \lambda_h \lambda_\Delta}\,.
\eea
From \eq{vevs}  we see that since $v_W=246$ GeV, $v_\Delta\sim$ keV may be obtained only by admitting  a fine tuning of order $\sim10^{-12}$. Moreover even by allowing such a fine tuning the breaking of the continuos lepton number gives rise to a Goldstone boson (GB), the so called Majoron, that  interacts with  neutrinos through  a coupling $m_\nu/v_\Delta \sim y_\nu m_D^2/M^2\sim10^{-3}$.  Bounds on  neutrino-Majoron coupling are obtained by no-observation of $\beta \beta$-decays and in pion and kaon decay experiments\cite{Bernatowicz:1992ma,Gelmini:1982rr,Lessa:2007up}. Nevertheless the strongest bounds are  obtained by analyzing supernova explosion\cite{Kachelriess:2000qc} and cosmic microwave background (CMB)\cite{Hannestad:2005ex}. All these analysis have been performed taking into account the 3 SM lepton generations but for our purposes we may take as reference value the bound indicated  for the diagonal couplings, $g\leq 10^{-7}$, being the off-diagonal ones even more restrictive. The present  bound is  four order of magnitude smaller than the value expected in the ISS formulation so far sketched, hence ruling it out. 

It is true that such a scheme may be saved avoiding  the problem of the massless Majoron   by   assuming that  lepton number is explicitly softly broken. Neverthless the naturalness problem would not be  solved yet  thus requiring looking for  alternative solutions.

Furthermore to be noticed that  in this scheme  we could not gauge  the global lepton number because   the new gauge boson  would  be  too  light and automatically excluded by LEP analysis\cite{PDG,Salvioni:2009mt}.

Let us  now suppose that lepton number is spontaneously broken by the VEV of a SM  singlet $\tilde{\Delta}$  with lepton charge $-1$  and  $\vev{\tilde{\Delta}} \sim v_W$. To implement the ISS we would need  an \emph{hidden} sector that gives rise to the effective operator
\beq
 y_S \frac{\tilde{\Delta}^2}{\Lambda_{eff}} S S \,,
\eeq
where $\Lambda_{eff}$ is an  effective scale that should be $\sim \vev{\tilde{\Delta}}^2/\mu\sim v_W^2/\mu\sim10^7 $ TeV. Clearly this operator may be originated by integrating out heavy fermions  with a mass around $10^7 $ TeV but this would shift the cutoff of our model to $\sim 10^8$ TeV thus reintroducing a new tension between   the TeV sterile neutrinos  scale  and this new cutoff. On the other hand this operator could be originated at the loop level: in this case we may write $1/\Lambda_{eff}$ as
\beq
\frac{1}{\Lambda_{eff}}= c\, \left(\frac{1}{16\pi^2}\right)^n \frac{1}{\Lambda}\,,
\eeq
where $n$ is the number of loops, $c$ summarizes the product of  different factors and couplings  that enter in the loops and $\Lambda$ may now be taken between 1 and $10$ TeV.   For $c\sim 0.1-1$  we need $n=2-3$ to sufficiently suppress the effective operator. 

In the following section we will show a minimal SM extension that implements this structure.

\section{The mechanism}

At the SM field content we add the ISS model  sterile fermion content, $\nu^c$ and $S$, and 3 new scalar SM singlets:  a real scalar field $\phi$, uncharged under the lepton number, and two complex fields  $\tilde{\Delta}$ and $\Delta$ with lepton charges $-1$ and $-2$ respectively. The Yukawa lagrangian involving the new fields coincides with the one given in \eq{ISS1}

\beq
\label{ISS1b}
\mathcal{L}= y_\nu\, L H \nu^c+ M \, \nu^c S + \frac{1}{2}\,y_S\Delta  \, S S\,+\frac{1}{2}\, y_{\nu^c} \Delta^\dag  \nu^c \nu^c\,,
\eeq
 while the  SM scalar potential is modified to
 \beq
 \label{scpot}
 V[h,\phi,\tilde{\Delta},\Delta]= V_h+ V_{sing}+V_{h sing}\,,
 \eeq
 where
  \bea
 &&V_h= \mu^2_h (h^\dag h)+ \lambda_h (h^\dag h)^2\,;\nn\\
  &&\nn\\
&& V_{sing}=  k \phi+ \mu^2_\phi \phi^2+ A_\phi \phi^3 +\lambda_\phi \phi^4 + \mu^2_{\tilde{\Delta}} (\tilde{\Delta}^\dag \tilde{\Delta})+\mu^2_\Delta (\Delta^\dag \Delta)+\nn\\
 &&\qquad+ A_{\tilde{\Delta}} \phi (\tilde{\Delta}^\dag \tilde{\Delta})+A_{\Delta} \phi ({\Delta}^\dag {\Delta})+ (\lambda e^{i \alpha}\tilde{\Delta}^{\dag 2}\Delta+H.c.)+ (B e^{i\beta} \phi \tilde{\Delta}^{\dag 2}\Delta+H.c.)\nn\\
 &+&  \lambda_{\tilde{\Delta}} (\tilde{\Delta}^\dag \tilde{\Delta})^2+\lambda_\Delta (\Delta^\dag \Delta)^2+ \lambda_{\tilde{\Delta}\phi}\phi^2 (\tilde{\Delta}^\dag \tilde{\Delta})+ \lambda_{\Delta \phi} \phi^2 (\Delta^\dag \Delta)+\lambda_{\Delta \tilde{\Delta}} (\Delta^\dag \Delta)(\tilde{\Delta}^\dag \tilde{\Delta})\,;\nn\\
&&\nn\\
&& V_{h sing}=A_h \phi (h^\dag h)+   \lambda_{h\phi}\phi^2 (h^\dag h)+  \lambda_{h\tilde{\Delta}} (h^\dag h)(\tilde{\Delta}^\dag \tilde{\Delta})+  \lambda_{h\Delta} (h^\dag h)(\Delta^\dag \Delta)\,.
 \eea
The mechanism we are proposing works if
\begin{itemize}
\item[1)] $\Delta$ is inert and does not develop a VEV;
\item[2)] all the neutral real components of the scalar fields mix. This mixing would induce the $\mu$ term given in \eq{mass1}  at the two-loop level.
\end{itemize}
Considering the first derivative system given by
\beq
\frac{\partial V[h,\phi,\tilde{\Delta},\Delta]}{\partial \varphi^\alpha_i}=0\,,
\eeq
where $\varphi=(h,\phi,\tilde{\Delta},\Delta)$ and $\alpha$ runs on all the scalar components of the field $\varphi_i$. The
  vacuum configuration given by
\bea
\label{vac}
\vev{h}=v_W/\sqrt{2} &\quad &\vev{\phi}=v_\phi \nn\\
\vev{\tilde{\Delta}}=v_{\tilde{\Delta}}/\sqrt{2}&\quad& \vev{\Delta}=0\,.
\eea
is a minimum of the scalar potential when
\bea
\beta&=&\alpha+\pi\,,\quad B=-\frac{\lambda}{v_\phi}\,,\nn\\
\mu^2_h&=&-\frac{1}{2}(2 A_{h} +2 \lambda_{h\phi} v_\phi^2+\lambda_{h\tilde{\Delta}} v_{\tilde{\Delta}}^2+ 2\lambda_h v_W^2)\,,\nn\\
\mu^2_\phi&=&-\frac{1}{4 v_\phi}( 2 k+ A_\phi v_W^2 + A_{\tilde{\Delta}} v_{\tilde{\Delta}}^2+2 A_\phi v_\phi^2+ 2 \lambda_{\tilde{\Delta}\phi}v_{\tilde{\Delta}}^2 v_\phi + 2 \lambda_{h \phi} v_W^2 v_\phi+2 \lambda_\phi v_\phi^3)\,, \nn\\
\mu^2_{\tilde{\Delta}}&=& -\frac{1}{2} ( 2 A_{\tilde{\Delta}} v_\phi+ 2\lambda_{\tilde{\Delta}\phi}v_\phi^2+2 \lambda_{\tilde{\Delta}\phi} v_\phi^2+ \lambda_{h\tilde{\Delta}}  v_W^2+ 2\lambda_{\tilde{\Delta}} v_{\tilde{\Delta}}^2)\,.
\eea
The imaginary part of $\tilde{\Delta}$ gives rise to the massless Majoron, while the imaginary part of $\Delta$ to a CP odd neutral state $a$ with mass
\beq
\label{masspse}
m_a^2=\mu^2_\Delta+A_\Delta v_\phi+ \lambda_{\Delta\phi} v_\phi^2+\frac{1}{2} \lambda_{\Delta \tilde{\Delta}} v_{\tilde{\Delta}}^2+\frac{1}{2}\lambda_{h\Delta} v_W^2\,.
\eeq
Among the 4 components of the SM higgs doublet, 3 correspond to the GBs eaten by the SM gauge bosons, while the neutral CP even component mixes with the neutral CP even components of the SM  singlets through
\bea
&&M^2_0=\nn\\
&& \left(\begin{array}{cccc} 2\lambda_h v_W^2 & A_h v_W + 2 \lambda_{h\phi} v_\phi v_W&\lambda_{h\tilde{\Delta}} v_W v_{\tilde{\Delta}}&0\\
A_h v_W + 2 \lambda_{h\phi} v_\phi v_W& -\frac{1}{2 v_\phi}( 2 k+ A_\phi v_W^2 + A_{\tilde{\Delta}} v_{\tilde{\Delta}}^2+2 A_\phi v_\phi^2+ 2 \lambda_{h \phi} v_W^2 v_\phi+4 \lambda_\phi v_\phi^3) &  A_{\tilde{\Delta}} v_{\tilde{\Delta}} + 2 \lambda_{\tilde{\Delta}\phi} v_\phi v_{\tilde{\Delta}}  &-\frac{\lambda \cos\alpha v_{\tilde{\Delta}}^2}{\sqrt{2}v_\phi} \\
\lambda_{h\tilde{\Delta}} v_W v_{\tilde{\Delta}}& A_{\tilde{\Delta}} v_{\tilde{\Delta}} + 2 \lambda_{\tilde{\Delta}\phi} v_\phi v_{\tilde{\Delta}}& 2 \lambda_{\tilde{\Delta}} v_{\tilde{\Delta}}^2 &0 \\
0& -\frac{\lambda \cos\alpha v_{\tilde{\Delta}}^2}{\sqrt{2}v_\phi} & 0& m_a^2 \end{array}\right)\,,\nn\\
&&
\eea
where $m_a^2$ is given in \eq{masspse} and the mass matrix $M^2_0$ is written in 
the basis $(h,\phi,\tilde{\Delta},\Delta)$. $M^2_0$ has a no vanishing  determinant, thus we do not have additional massless particles.

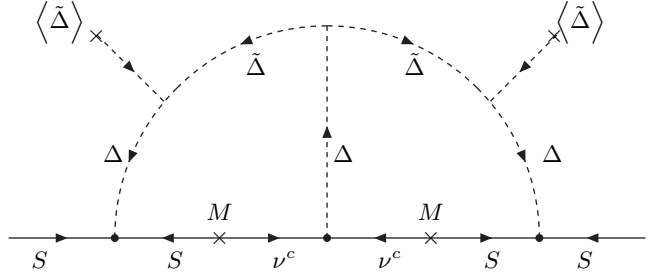
\begin{figure}
\begin{picture}(300,90)(0,0)
\Vertex(90,15){1.5}
\Vertex(250,15){1.5}
\Vertex(170,15){1.5}
 \DashArrowArc(170,15)(80,135,180){2}
  \DashArrowArc(170,15)(80,90,135){2}
\DashArrowArcn(170,15)(80,45,0){2}
\DashArrowArcn(170,15)(80,90,45){2}
 \DashArrowLine(170,15)(170,95){2}
 \DashArrowLine(83.33,91.56)(108.44,66.56){2}
  \DashArrowLine(256.56,91.56)(231.56,66.56){2}
  \Text(83.33,91.56)[c]{$\times$}
\Text(256.56,91.56)[c]{$\times$}
\Text(130,15)[c]{$\times$}
\Text(210,15)[c]{$\times$}
\Text(130,25)[c]{$M$}
\Text(210,25)[c]{$M$}
\ArrowLine(210,15)(170,15)
\ArrowLine(210,15)(250,15)
 \ArrowLine(290,15)(250,15)
\ArrowLine(130,15)(170,15)
\ArrowLine(130,15)(90,15)
\ArrowLine(50,15)(90,15)
\Text(65,10)[rt]{$S$}
\Text(265,10)[l t]{$S$}
\Text(110,10)[l t]{$S$}
\Text(150,10)[l t]{$\nu^c$}
\Text(190,10)[l t]{$\nu^c$}
\Text(230,10)[l t]{$S$}
\Text(86,50)[l t]{$\Delta$}
\Text(252,50)[l t]{$\Delta$}
\Text(173,50)[l t]{$\Delta$}
\Text(60,105)[l t]{$\vev{\tilde{\Delta}}$}
\Text(140,85)[l t]{$\tilde{\Delta}$}
\Text(200,85)[l t]{$\tilde{\Delta}$}
\Text(257,105)[l t]{$\vev{\tilde{\Delta}}$}
\end{picture}
 \caption{The origin of the $\mu$ term that gives rise to neutrino masses through \eq{mass1}. In a similar way the $\tilde{\mu}$ term is generated, but  its effect is subleading as explained in sec.~\ref{tow}.} \label{loops}
\end{figure}

The $\mu$ term is  induced at the two-loop level thanks to the mixing of the 4 neutral CP even states as may be seen in fig.~\ref{loops}. The presence of the  singlet $\phi$ is fundamental in order to allow the vacuum configuration given in \eq{vac}: without $\phi$, $\Delta$ could not behave as an inert scalar thus destroying the full mechanism.

The $\mu$ term expression  is roughly given
\beq
\label{loopmu}
\mu\sim y_S^2 y_{\nu^c} \frac{1}{(16 \pi^2)^2} A^3\frac{v_{\tilde{\Delta}}^2}{M^4 }\sim y_S^2 y_{\nu^c} 10^{-5} \frac{v_W^5}{(1 \,\mbox{TeV})^4}\sim y_S^2 y_{\nu^c} 10^{-5}\, \mbox{GeV}\sim y_S^2 y_{\nu^c} 10\,\mbox{keV}\sim 1\,\mbox{keV}\,,
\eeq
where $A^3\sim v_W^3$ stays for the product  of scalar potential trilinear couplings and we have assumed $y_S, y_{\nu^c}<1$.

To be noticed that in this scheme the neutrino-Majoron coupling  $y_J$  is sufficiently suppressed to satisfy   all the constraints\cite{Kachelriess:2000qc,Hannestad:2005ex}
\beq
y_J\sim \frac{m_\nu}{v_{\tilde{\Delta}}}\sim 10^{-11}\,.
\eeq

\section{Outlook: Gauged $B-L$ number in the dynamical ISS scenario}

In the previous section we have provided a mechanism that furnishes a justification for the keV scale of the ISS model.  Furthermore  the associated Majoron is sufficiently weakly coupled to neutrinos not to be ruled out by the most recent analysis.

Nevertheless we may ask what could change in the gauged version of the mechanism proposed. In this section we will briefly sketched the main features of the gauge version of the model so far described, leaving for a future work the detailed analysis.

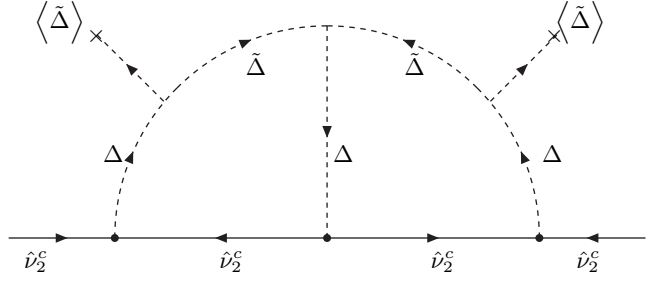
\begin{figure}
\begin{picture}(300,90)(0,0)
\Vertex(90,15){1.5}
\Vertex(250,15){1.5}
\Vertex(170,15){1.5}
 \DashArrowArcn(170,15)(80,180,135){2}
  \DashArrowArcn(170,15)(80,135,90){2}
\DashArrowArc(170,15)(80,0,45){2}
\DashArrowArc(170,15)(80,45,90){2}
 \DashArrowLine(170,95)(170,15){2}
 \DashArrowLine(108.44,66.56)(83.33,91.56){2}
  \DashArrowLine(231.56,66.56)(256.56,91.56){2}
  \Text(83.33,91.56)[c]{$\times$}
\Text(256.56,91.56)[c]{$\times$}
\ArrowLine(170,15)(250,15)
 \ArrowLine(290,15)(250,15)
\ArrowLine(170,15)(90,15)
\ArrowLine(50,15)(90,15)
\Text(65,10)[rt]{$\hat{\nu}^c_2$}
\Text(265,10)[l t]{$\hat{\nu}^c_2$}
\Text(130,10)[l t]{$\hat{\nu}^c_2$}
\Text(210,10)[l t]{$\hat{\nu}^c_2$}
\Text(86,50)[l t]{$\Delta$}
\Text(252,50)[l t]{$\Delta$}
\Text(173,50)[l t]{$\Delta$}
\Text(60,105)[l t]{$\vev{\tilde{\Delta}}$}
\Text(140,85)[l t]{$\tilde{\Delta}$}
\Text(200,85)[l t]{$\tilde{\Delta}$}
\Text(257,105)[l t]{$\vev{\tilde{\Delta}}$}
\end{picture}
\caption{Radiative correction that induces the $\hat{\nu}^c_2$ mass $\sim $  MeV.} \label{loopsb}
\end{figure}

Once  the lepton number is gauged, in order to erase the triangle anomalies of the kind $SU(2)-SU(2)-U(1)_L$ and  $U(1)_Y-U(1)_Y-U(1)_L$ where $SU(2)$ and $U(1)_Y$ are the EW SM symmetries and $U(1)_L$ the  lepton symmetry, we are forced gauging the global symmetry $U(1)_{B-L}$ and not only $U(1)_L$. So far nothing changes for  what concerns neutrino masses and the scalar potential discussion. However, due to the presence of the singlet $S$,   a triangle anomaly is still left, that related to the triangle   $U(1)_{B-L}-U(1)_{B-L}-U(1)_{B-L}$.  In our ISS scheme we may get an anomaly free $U(1)_{B-L}$  by adding 2 right-handed neutrinos, $\nu^c_1$ and $\nu^c_2$, per each $S$ singlet.  In this case the  neutrino Yukawa lagrangian  in \eq{ISS1b}  becomes

\beq
\label{ISS2}
\mathcal{L}= y_{\nu_1}\, L H \nu_1^c+ y_{\nu_2}\, L H \nu_2^c + M_1 \, \nu^c_1 S + M_2 \, \nu^c_2 S+ 1/2 y_S\Delta  \, S S\,+ 1/2 y^a_{\nu^c_1} \Delta^\dag  \nu_1^c \nu_1^c+ 1/2 y^a_{\nu^c_2} \Delta^\dag  \nu_2^c \nu_2^c+y^b_{\nu^c} \Delta^\dag  \nu_1^c \nu_2^c+ H.c.\,.
\eeq

and  after EW and lepton number spontaneous breaking the neutrino mass matrix  given  in \eq{mass1} turns into
\beq
\label{mass2}
M_\nu= \left ( \begin{array}{cccc} 0&m_{D_1} &m_{D_2} &0\\ m_{D_1} &\tilde{\mu}_1 & \tilde{m}&M_1\\   m_{D_2} &\tilde{m} & \tilde{\mu}_2&M_2\\0 &M_1& M_2& \mu \end{array}\right)\sim  \left ( \begin{array}{cccc} 0&m_{D_1} &m_{D_2} &0\\ m_{D_1} &0 &0&M_1\\   m_{D_2} &0& 0&M_2\\0 &M_1& M_2& 0 \end{array}\right)\,,
\eeq
since $\mu,\tilde{\mu}_{1,2}, \tilde{m}\ll m_{D_{1,2}}\ll M$\footnote{$\mu,\tilde{\mu}_{1,2}, \tilde{m}$ would be all induced at the two-loop level.}. To prevent  that left handed neutrinos participate to a  GeV scale Dirac  neutrino we need to impose a permutation matter symmetry  between $\nu^c_1$ and $\nu^c_2$.  In this  way  $y_{\nu_1}=y_{\nu_2}=y_{\nu}$, $M_1=M_2=M$ and $y^a_{\nu^c_1}=y^a_{\nu^c_2}=y^a_{\nu^c}$.
Defining the right-handed neutrino basis
\bea
\hat{\nu}^c_1&=& \frac{1}{\sqrt{2}} (\nu_1^c+\nu_2^c)\,,\nn\\
\hat{\nu}^c_2&=& \frac{1}{\sqrt{2}} (-\nu_1^c+\nu_2^c)\,,
\eea
 \eq{ISS2} becomes
\beq
\label{ISS3}
\mathcal{L}= y_\nu\,\sqrt{2}\, L H \hat{\nu}_1^c  + \sqrt{2} M \, \hat{\nu}^c_1 S + 1/2 y_S\Delta  \, S S\,+ 1/2(y^a_{\nu^c}+y^b_{\nu^c})\Delta^\dag  \hat{\nu}_1^c  \hat{\nu}_1^c+ 1/2(y^a_{\nu^c}-y^b_{\nu^c})\Delta^\dag  \hat{\nu}_2^c  \hat{\nu}_2^c +H.c.\,.
\eeq
Clearly  in this basis the permutation symmetry is now a $Z_2$--matter--symmetry under which $\hat{\nu}^c_2$ is odd while $\hat{\nu}^c_1$ is even as all the other fermion and scalar particles. 
Light neutrino masses are  induced as in the previous formulation and only $\hat{\nu}^c_1$ participates in the ISS mechanism. On the other hand $\hat{\nu}^c_2$ is a sterile neutrino decoupled by all the other fermions whose mass is generated at the two-loop level as it happens for the $\mu$ term. However the radiative correction for $\hat{m}_2 $ is slightly different with respect to the one that gives rise to $\mu$  as may be seen by  looking at fig.~\ref{loopsb}:  $\hat{\nu}^c_2$ does not   participate in a quasi-Dirac spinor and therefore only $\hat{\nu}^c_2$ runs in the loops. As consequence  
\beq
\hat{m}_2 \sim    \hat{y}_{\nu}^3 \frac{1}{(16 \pi^2)^2}\frac{A^3}{m_S^2 }\sim \hat{y}_{\nu}^310^{-5} {v_W}\sim \,0.1-1\,\mbox{MeV}\,,
\eeq
where $A^3\sim v_W^3$ as in \eq{loopmu}, $m_S$ stays for the generic  scalar mass of the fields in the loops and $\hat{y}_{\nu}=1/2(y^a_{\nu^c}-y^b_{\nu^c})$.

In this case we have found out a very nice link between the $\mu$ term in the ISS model and the presence of a MeV sterile neutrino.  This sterile neutrino interacts weakly with the new gauge boson $Z'$ and with the new singlet sector through its coupling with $\Delta$  and it is  stable because of the $Z_2$ symmetry, thus it is a possible MeV DM candidate. The scenario underlined is similar to that proposed in \cite{Hooper:2008im}, even if  in that case the model was supersymmetric.  The full analysis of this candidate as  plausible DM would be addressed in a following paper\cite{iofut}.

The phenomenology of the model in its gauged formulation is much more interesting: 
 the new  $Z'$ may be produced at LHC  and then detected  through the usual $U(1)_{B-L} $ channels\cite{Salvioni:2009mt,Emam:2007dy} or through its decays into the new neutral scalar sector, thus giving rise to specific signatures. Even these aspects will be addressed afterwards\cite{iofut}.

\section{Conclusion}

The ISS mechanism is one of the most appealing mechanism introduced to explain neutrino masses: lepton number is almost an approximate symmetry of the lagrangian, being broken by a very small mass term in a new fermion sector.  The presence of the lepton number breaking parameter induces tiny left handed neutrino masses, that vanish when the lepton symmetry is restored. Albeit one may invoke t'Hooft's criteria~\cite{'tHooft:1979bh}   to justify the smallness of $\mu$, notwithstanding 
the original ISS model  does not provide any explanation of it. 

In this paper we have proposed a minimal model in which the two unity  lepton number breaking parameter is induced at the two-loop level. The model has been built  following the two criteria of \emph{naturalness} and \emph{minimality}.  The new physics scale is around the TeV and the lepton number breaking scale is comparable to the EW one. The simplest version of the model has a reduced number of  new degrees of freedom with respect to the usual  ISS model:  three SM scalar singlets, $\phi$, $\tilde{\Delta}$ and ${\Delta}$ with lepton charges $0,-1$ and $-2 $ respectively.  ${\Delta}$ is inert while $\phi$ and $\tilde{\Delta}$ develop a VEV around the EW scale. We have checked that the vacuum configuration proposed is indeed a minimum of the scalar potential. Moreover the presence of $\phi$ is needful  because it allows keeping $\Delta$ inert while $\tilde{\Delta}$ develops a VEV and it provides their mixing. Thanks to this the loop in fig.~\ref{loops}  gives a no null contribution. 

The gauged version  of the model proposed is slightly less minimal but much more phenomenologically interesting: by requiring anomaly cancellation and neutrino masses preservation we get a MeV sterile neutrino stable under a matter $Z_2$ symmetry that could be a good DM candidate. Moreover the presence of the new gauge boson $Z'$ and of the new neutral scalars below the TeV scale gives rise to testable  and characteristic signatures at the LHC that may be studied in a following   paper.

\section*{Aknowledgments}

The author  thanks S. Morisi and M. Lattanzi for very useful comments and suggestions.
This work  is part of the research program of the Foundation for Fundamental Research of Matter (FOM) and it  has also been partially supported by the National Organization for Scientific Research (NWO).

\bibliographystyle{h-physrev4}

\end{document}